\documentclass[11pt]{article}

\usepackage[fleqn]{amstex} 
\usepackage{amssymb}
\usepackage{latexsym}
\usepackage{epsfig}   
\usepackage{subfigure}
\usepackage{graphicx}

\newcounter{no}

\newcommand{\vect}[1]{\boldsymbol{#1}}

\newcommand{\noi}{\noindent}

\newcommand{\eq}[1]{\begin{equation} #1 \end{equation}}

\newcommand{\set}[1]{\{ #1 \}}
\def\R{{\rm I\!R}} 		
\def\Z{{\mathbb Z}}
\newcommand{\acroix}{a^{\dagger}}
\newcommand{\id}{{}^{\phantom{\dagger}}}
\newcommand{\ie}{\emph{i.e.\ }}
\newcommand{\circuit}{{\cal} C}
\newcommand{\sig}{\sigma}

\begin{document}

\title{Falicov-Kimball Models: A Partial Review of the Ground States Problem}
\author{Christian Gruber \\
{\em Institut de Physique Th\'{e}orique} \\
{\em Ecole Polytechnique F\'{e}d\'{e}rale de Lausanne} \\
{\em CH-1015 Lausanne, Switzerland } \\
{\em Christian.Gruber\@epfl.ch}
}
\maketitle
\begin{abstract}
In this review we present a biased review of the ground 
state properties of the Falicov-Kimball models in 1, 2 and $\infty$ 
dimensions, considering either fermions or hard core bosons.  In particular 
we want to show the very rich structure that these models exhibit, and 
to point out the analogies and differences associated with the 
statistic of the quantum particles and the nature of the lattice 
(bipartite or not). The flux phase problem is also discussed.
\end{abstract}

\vspace{1 cm}
\noi {\large \bf Contents}: \hfill \\ \vspace{-0.5cm}
\begin{enumerate}
\item Introduction \\
    1.1 Definition of the models\\
    1.2 Interest of the Falicov-Kimball model \\
    1.3 Problems \\
    1.4 Methods
\item Results for 1-dimensional systems: ground states, Peierls 
instability, conjecture
\item Results for 2-dimensional systems: ground states and flux phases 
on the square and triangular lattices
\item Preliminary results for $\infty$-dimensions Bethe lattice
\item Conclusions
\item References
\end{enumerate}
\nocite{*}

\section{Introduction}
\subsection{Definition of the model}
The Falicov-Kimball model is a \emph{lattice model} of \emph{quantum 
particules} interacting with \emph{classical particles.}

The \emph{lattice} $\Lambda=\set{x}$ is defined by a finite or 
infinite set of sites $x$ in $\R^{d}$. Usually one considers 
$d$-dimensional cubic lattices $\Z^{d}$, and $\Lambda \subset 
\Z^{d}$. However it is interesting to consider more general lattices, 
e.g. the 2-$d$ triangular lattice which, contrary to cubic lattices, 
is not bipartite. Indeed, particles on bipartite lattices have 
particle-hole symmetries, and one would like to know what properties 
remain valid for systems which do not have these symmetries. As we 
shall review, the rigourous results which have been obtained so far 
concern the cases where the coupling between quantum and classical 
particles is either very strong (in any dimension), or very weak (for 
$d$=1). For intermediate coupling, the only results concern reduced phase diagrams (sec. 1.4) and have been obtained using exact numerical methods. To investigate this 
intermediate coupling regime, it appears of interest to study first 
the model on the Bethe lattice in the limit where the coordination 
number becomes infinite; indeed in this limit one obtains a mean 
field model which should be simpler to study rigorously.

In the original model, the \emph{quantum particles}, called 
``electrons'' in the following, are spinless fermions described by 
creation and annihilation operators $\acroix_{x}, a\id_{x}$, 
satisfying the usual anticommutation relations
\eq{
    \set{a\id_{x},a\id_{y}} = \set{\acroix_{x},\acroix_{y}} = 0,
    \qquad \set{\acroix_{x},a\id_{y}} = \delta_{x,y}.
}
It is well known that the Pauli principle is important to get a 
crystalline state, since such a state can not occur if fermions are 
replaced by bosons \cite{4}. However hard-core  bosons do indeed lead 
to crystaline structures \cite{1} and they are interesting to study. 
Spinless hard-core bosons are described by operators $\acroix_{x}, 
a\id_{x}$ satisfying the following relations
\eq{
    \set{\acroix_{x},a\id_{x}} = 1, \qquad 
\set{\acroix_{x},\acroix_{x}} = 
    	\set{a\id_{x},a\id_{x}} = 0
}
but
\eq{
    [a_{x}^{\#},a_{y}^{\#}] = 0 	\quad \text{if} \quad x \neq y
    	\qquad (a_{x}^{\#} = \acroix_{x}, a\id_{x})
}

For the finite system $\Lambda$, the \emph{kinetic energy} of the 
quantum particles is defined by the operator
\eq{
    K_{\Lambda} = -\sum_{x,y \in \Lambda} t_{xy} \, \acroix_{x} 
a\id_{y}
}
where $t_{xy} = t_{yx}^{*} = |t_{xy}| e^{i\theta_{xy}}$.
Complex hopping constants $t_{xy}$ are introduced to model particles 
in an external magnetic field: given a circuit ${\circuit} = 
(x_{1},x_{2}, \ldots, x_{n}, x_{n+1}=x_{1})$, \ie an ordered sequence 
of sites $x_{i}$ in $\Lambda$ such that $t_{x_{i},x_{i+1}} \neq 0$ 
for all $i=1, \ldots, n$, the magnetic flux through this circuit is 
defined by
\eq{
    \phi_{\circuit} =  \sum_{i=1}^{n} \theta_{x_{i},x_{i+1}} \mod 
2\pi.
}
In the original Falicov-Kimball model, only hopping between 
nearest-neighbour sites were considered, with
\eq{
    t_{xy} = \begin{cases}
    	t \in \R	& \text{if\ } |x-y| = 1	\\
    	0	& \text{otherwise.}
    	\end{cases}
}
Since then, several extensions of the model have been investigated, 
introducing quantum particles with ``spin'', described by operators 
$\acroix_{x\sig},a\id_{x\sig}$, together with the kinetic energy
\eq{
    K_{\Lambda} = - \sum_{x,y \in \Lambda} \,\sum_{\sig,\sig'} 
    	t_{xy}^{\sig\sig'} \, \acroix_{x\sig} a\id_{y\sig'}.
}
For example, in the standard Hubbard model
\eq{
    K_{\Lambda} = -\sum_{x,y \in \Lambda}\, \sum_{\sig = 
\uparrow,\downarrow}
    	t_{xy} \, \acroix_{x\sig} a\id_{y\sig}.
}
For the asymmetric Hubbard model
\eq{
    K_{\Lambda} = - \sum_{x,y \in \Lambda} \,\sum_{\sig = 
\uparrow,\downarrow}
	t_{xy}^{\sig} \, \acroix_{x\sig} a\id_{y\sig}
}
with the Falicov-Kimball model given by
\eq{
    t_{xy}^{\uparrow} = t_{xy}, \quad t_{xy}^{\downarrow} = 0.
}
For the Montorsi-Rasetti model
\eq{
    K_{\Lambda} = -\sum_{x,y \in \Lambda} \sum_{\sig,\sig' = 
\uparrow,\downarrow}
	t_{xy} \, \acroix_{x\sig} a\id_{y\sig'}
}
and it has been shown that this model is equivalent to the 
Falicov-Kimball model.

The \emph{classical particles}, called ``ions'' in the following, are 
assumed to have a hard-core. They are described by random variables 
$W_{x}\in\set{0,1}, x\in\Lambda$, where $W_{x}=1$ means that the site 
$x$ is occupied by an ion and $W_{x}=0$ means that the site $x$ is 
not occupied by an ion. In the spin-language, the random variable is 
$s_{x}$, where $s_{x} = 2W_{x}-1 \in \set{-1,+1}$.
The classical particles could be ions, impurities, spins, localised 
f-electrons, ..., depending on the physical system under 
investigation.

Again several extension have been considered. For example in the 
static Holstein model $s_{x} \in\R$ (and is associated with phonons), 
while in the static Kondo model 
$\vect{s}_{x}=(s_{x}^{1},s_{x}^{2},s_{x}^{3}) \in \R, 
|\vect{s}_{x}|=1$, describes localized magnetic impurities.

The classical particles do not move and they do not have any kinetic 
energy. However in the definition of equilibrium states one will take 
annealed averages over all possible configurations of ions.

Except for the hard-core between ions, and the statistics for 
electrons, it is assumed that the only \emph{interactions} are 
between electrons and ions when they occupy the same lattice site. It 
models for example the screened Coulomb interaction $U$ which can be 
either positive (repulsion), ore negative (attraction). In other 
words, given the configurations $\vect{W} = 
\set{W_{x}}_{x\in\Lambda}$, or $\vect{s} = 
\set{s_{x}}_{x\in\Lambda}$, of the ions, the interaction is described 
by the operator
\newcommand{\inter}{\text{int}}
\eq{
    H_{\Lambda}^{\inter}(\vect{W}) = 2 U \sum_{x\in\Lambda}
    	\left(W_{x}-\frac{1}{2}\right)
    	\left(\acroix_{x}a\id_{x} - \frac{1}{2}\right)
}
or, equivalently,
\eq{
    H^{\inter}(\vect{s}) = U \sum_{x\in\Lambda} s_{x} \left(  
\acroix_{x}a\id_{x} - \frac{1}{2} \right).
}
The factor $1/2$ is introduced for convenience to exhibit symmetry 
properties. In the canonical formalism it amounts simply to adding a 
constant; in the grand canonical formalism to a redefinition of the 
chemical potentials.
In conclusion, given a configuration of the ions, the hamiltonian of 
the finite system $\Lambda$ is given by
\eq{
    H_{\Lambda}(\vect{s}) = K_{\Lambda} + 
H_{\Lambda}^{\inter}(\vect{s})
}
to which is added in the grand canonical formalism the contribution
\eq{
    -\mu_{e}\sum_{x\in\Lambda} \acroix_{x}a\id_{x} - 
\mu_{i}\frac{1}{2}\sum_{x\in\Lambda} (s_{x} + 1 ),
}
with $\mu_{e},\mu_{i}$ the chemical potentials for electrons and ions.

Again several extensions can be considered. For example, for 
electrons with spins the following hamiltonian has been considered
\eq{
    H_{\Lambda}(\vect{s}) = K_{\Lambda} + U \sum_{x\in\Lambda} s_{x} 
\left(n_{x}-\frac{1}{2}\right) + U' \sum_{x\in \Lambda} \left( 
\acroix_{x\uparrow}a\id_{x\uparrow} - \frac{1}{2}\right) \left( 
\acroix_{x\downarrow}a\id_{x\downarrow} -\frac{1}{2} \right)
}
where $n_{x} = \acroix_{x\uparrow}a\id_{x\uparrow} + 
\acroix_{x\downarrow}a\id_{x\downarrow}$.

\subsection{Interest of the Falicov-Kimball model}

One of the most fundamental problem in condensed matter physics is to 
understand the phenomenon of phase transitions, in particular why all 
elements and many compounds crystallize in periodic structures. It is 
well known that one of the driving principles behind this ordering is 
associated with the structure of the ground states; however it is 
still not clear what physical mechanisms are necessary for phase 
transitions to occur, especially for quantum systems. Indeed, because 
of the quantum fluctuations, the ground state properties, as well as 
the low temperature behavior, are difficult to extract for quantum 
systems.

In 1969 Falicov and Kimball \cite{5} introduced their model to study 
metal-insulator transitions in mixed valence compounds of rare earth 
and transition metal oxides as an effect of the interaction between 
localised f-electrons (``classical particles'') and itinerant 
d-electrons (quantum particles). Later this same model was considered 
to study ordering in mixed valence systems, order-disorder 
transitions in binary alloys, and itinerant magnetism.

This model was reinvented in 1986 by Kennedy and Lieb \cite{4} as a primitive 
model for matter to study crystallisation. In this interpretation 
ions are represented by classical particles and itinerant electrons 
by quantum particles. Depending on whether there is one, or several, 
electronic bands near the Fermi level, one is lead to consider 
neutral (equal density of ions and electrons), or non-neutral 
systems. For $\mu_{e}=\mu_{i}=0$, which is the symmetry point of the 
system and corresponds to the neutral case with 
$\rho_{e}=\rho_{i}=\frac{1}{2}$, Kennedy and Lieb proved the 
existence of long range order (crystal) at low teperatures for any 
coupling $U$ and any dimension $d\geq 2$, together with the absence 
of such order (fluid) at high temperatures. Their result was then 
extenden by J. L. Lebowitz and N. Macris in 1994 \cite{6}, for values of 
chemical potential in a neighborhood of the symmetry point.

\subsection{Problems}

The first step to study phase transitions is to obtain the zero 
temperature phase diagram, \ie the ground states. \\
\rule[0.8cm]{0cm}{0cm}
\begin{minipage}[4cm]{\textwidth}
\noi \underline{Problem 1}: ``Ground states'' \ \ \ ($T=0$)

\hspace{0.8cm} Find the configuration of ions which minimize the 
energy.
\end{minipage} \vspace{0.0cm}

\noi Two approaches have been used to solve this problem.

In the \emph{canonical formalism}, given $(N_{e},N_{i})$, the number 
of electrons and ions, the problem is to find the configurations of 
ions $\vect{s}=\set{s_{x}}$ which minimize the energy, \ie
\eq{
    E_{N_{e}}(\vect{s}) = \min_{\vect{s}':\  N_{i}(\vect{s}')=N_{i}}
    	E_{N_{e}}(\vect{s}')
}
where $N_{i}(\vect{s})=\frac{1}{2}\sum_{x\in\Lambda}(s_{x}+1)$ and 
$E_{N_{e}}(\vect{s}')$ is the infimum of the spectrum of the 
hamiltonian $H_{\Lambda,N_{e}}(\vect{s}')$ restricted to the $N_{e}$ 
electrons sector.

\newcommand{\Tr}{\mathrm{Tr}}
\newcommand{\tend}{\rightarrow}
In the \emph{grand canonical formalism}, given $(\mu_{e},\mu_{i})$, 
the electron and ion chemical potentials, one starts from the 
partition function. Introducing an effective free energy $\cal F$ by 
means of a partial trace over the electrons degrees of freedom, \ie
\begin{equation}
    Z_{\Lambda}(\beta,\mu_{e},\mu_{i})
     = \sum_{\vect{s}} \Tr\set{
    	e^{-\beta\left[  H_{\Lambda}(\vect{s}) - \mu_{e}N_{e} -
    	\mu_{i}N_{i}(\vect{s})\right]}}
    = \sum_{\vect{s}} e^{-\beta {\cal 
F}_{\Lambda}(\vect{s};\beta,\mu_{e},\mu_{i})}
\end{equation}
one is lead to study a classical spin lattice system. Defining the 
effective hamiltonian for this classical spin system by
\eq{
    E_{\Lambda}(\vect{s}; \mu_{e},\mu_{i}) = \lim_{\beta\tend\infty} 
{\cal F}_{\Lambda}(\vect{s};\beta,\mu_{e},\mu_{i})
}
the problem is to find those configurations of ions $\vect{s}$ which 
minimize $E_{\Lambda}(\vect{s}';\mu_{e},\mu_{i})$, \ie to find the 
zero temperature phase diagram of the effective hamiltonian.

Since the interaction is on-site, the quantum-mechanical problem for 
fermions can be solved by looking at the eigenvalues 
$e_{j}(\vect{s})$ of the hamiltonian $h_{\Lambda}(\vect{s})$ for one 
electron moving in the potential $U(s_{x})$ defined by the 
configuration of ions. Therefore for \emph{fermions systems}
\begin{align}	\label{20}
    E_{\Lambda}(\vect{s}_{i};\mu_{e},\mu_{i})
    & = \sum_{e_{j}(\vect{s}) \leq \mu_{e}}
    	[e_{j}(\vect{s})-\mu_{e}] - \frac{U}{2}\sum_{x}s_{x} - 
\mu_{i}N_{i}(\vect{s}) \\
    &= E_{N_{e}}(\vect{s}) - \mu_{e}N_{e} -\mu_{i} N_{i}(\vect{s})
    	\qquad \text{with $N_{e} = N_{e}(\mu_{e})$.}
\end{align}

On the other hand for systems of bosons with hard cores, we have a 
truly many-body problem of interacting particles; in this case one 
has to proceed via the effective free energy and the limit $\beta 
\tend \infty$.

\noi \begin{minipage}[0cm]{\textwidth}
\rule[0.8cm]{0cm}{0cm} \noi \underline{Problem 2}: ``Flux phases'' \ 
\ \ ($T=0$)

\hspace{0.5cm}\parbox[1cm]{14.3cm}{Let $\Phi=\set{\phi_{\circuit}}$, 
where $\phi_{\circuit}$ denotes the magnetic flux through the 
elementary circuit~$\circuit$. The problems are the following:\\
i) Given $(\mu_{e},\mu_{i},\Phi)$, find the configurations of ions 
$\vect{s}$ which minimize 
$H_{\Lambda}(\vect{s};\Phi,\mu_{e},\mu_{i})$. \\
ii) Given $(\mu_{e},\mu_{i})$, find the configurations $\vect{s}$ and 
the fluxes $\Phi$ which minimize 
$H_{\Lambda}(\vect{s};\Phi,\mu_{e},\mu_{i})$.}
\end{minipage} \vspace{0.2cm}

\noi \begin{minipage}[0cm]{\textwidth}
\rule[0.8cm]{0cm}{0cm} \noi \underline{Problem 3}: ``Low themperature 
phase diagram'' \\

\vspace{-0.3cm}\hspace{0.5cm}\parbox[1cm]{14.3cm}{The real problem in 
the study of phase transitions is to show that the zero temperature 
phase diagram is stable at low temperatures.}
\end{minipage} \vspace{0.2cm}

\noi This question is discussed in the talks by R. Kotecky, D. 
Ueltschi, and N. Datta, and will not be considered in this review (see ref. [7]-[13]).

\subsection{Methods}

Several methods have been introduced to study the zero temperature 
phase diagram.

\paragraph{a) Expansion of $E_{\Lambda}(\vect{s};\mu_{e},\mu_{i})$ in 
powers of $|U|^{-1}$}  \hfill\\

This method has been introduced in both the canonical and the grand 
canonical formalism. It is valid in any dimension, but is restricted 
to
\eq{ \label{1}
    |U| > ct\quad\text{and} \quad \mu_{e} \in ]-|U|+ct,|U|-ct[
}
where $c$ is some constant depending on the lattice and $t = 
\max|t_{xy}|$.
The conditions \eqref{1} imply in particular that this technique is 
restricted to
\begin{alignat}{2} \label{23}
    \text{``neutral'' systems, \ie\ } &\rho_{e}=\rho_{i},&\quad 
\text{if\ } U<0 \\ \label{24}
    \text{``half-filled'' systems, \ie\ } &\rho_{e}+\rho_{i}=1,&\quad 
\text{if\ } U>0
\end{alignat}
where $\rho_{e}=\frac{N_{e}}{|\Lambda|}$ and 
$\rho_{i}=\frac{N_{i}}{|\Lambda|}$, are the (average) electrons and ions 
densities.

In the case of \emph{fermions}, the expansion of 
$E_{N_{e}}(\vect{s})$ in powers of $|U|^{-1}$ is easily obtained 
using \eqref{20}, together with the following property: with $z$ the 
maximal coordination number of the lattice and $|U|>zt$, then, for 
any ions configuration $\vect{s}$, the spectrum of the 1-electron hamiltonian $h_{\Lambda}(\vect{s})$ has a gap containing the interval $]-|U|+zt,|U|-zt[$; for $\mu_{e}$ inside this gap the number of 
eigenvalues $e_{j}(\vect{s})\leq \mu_{e}$ is $N_{i}(\vect{s})$ if 
$U<0$ and $|\Lambda|-N_{i}(\vect{s})$ if $U>0$.

\newcommand{\sss}{\vect{s}}
\newcommand{\romd}{\mathrm{d}}
\newcommand{\FF}{{\cal F}}
Therefore, for $N_{e}=N_{i}(\vect{s})$ if $U<0$, and for 
$N_{e}=|\Lambda|-N_{i}(\vect{s})$ if $U>0$, \ie for $\mu_{e}$ in the 
above gap, we have
\eq{
    E_{N_{e}}(\vect{s}) = -\frac{U}{2} \sum_{x\in\Lambda} s_{x} + 
\frac{1}{2\pi i} \oint_{\circuit} \romd z \, \Tr 
\left\{\frac{z}{z-h_{\Lambda}(\sss)} \right\}
}
where $\circuit$ is a contour in the complex plane enclosing all 
negative eigenvalues.

Iterating the resolvent identiy for $[z-h_{\Lambda}(\sss)]^{-1}$ we 
obtain explicitely the desired expansion
\eq{
    E_{N_{e}}(\sss) = \sum_{n\geq 1} \frac{1}{|U|^{n}} E_{n}(\sss)
}
with
\eq{
    E_{n}(\sss) = \sum_{x_{1},\ldots,x_{n+1}\in\Lambda} 
\frac{(-1)^{m}}{m}
    	\frac{(n-1)!}{(m-1)!(n-m)!} \prod_{i=1}^{n+1} t_{x_{i}x_{i+1}}
}
where the sequence $(x_{1},\ldots,x_{n+1})$ must contain at least one 
emply site and one occupied site, and $m=m(\sss)$ is the number of 
sites $x_{i}, i=1,\ldots,n+1$, such that $s_{x_{i}} = -1$ if $U>0$, 
and $s_{x_{i}}=+1$ if $U<0$.

For \emph{hardcore bosons}, the $|U|^{-1}$ expansion is obtained 
using the closed loop expansion of Messager-Miracle \cite{12} for the 
effective free energy $\FF(\beta;\sss,\mu_{e},\mu_{i})$. Taking the 
limit $\beta\tend\infty$ yields the expansion for 
$E_{\Lambda}(\sss,\mu_{e},\mu_{i})$ which is convergent if the 
conditions \eqref{1} are satisfied, and implies the restriction Eq. 
\eqref{23}, and \eqref{24}.

\newcommand{\smm}{\sss;\mu_{e},\mu_{i}}
In conclusion for both statistics we are able to write the effective 
hamiltonian as
\eq{
    E_{\Lambda}(\smm) = H^{(k)}(\smm) + R^{(k)}(\smm)
}
where
\eq{
    H^{(k)}(\smm) = H^{(0)}(\smm) + \sum_{n\geq 1}^{k} 
\frac{1}{|U|^{n}} E_{n}(\sss).
}

At this point, the strategy is to study the ground states of the 
truncated hamiltonian $H^{(k)}$ and to control the rest $R^{(k)}$. 
With this technique one can prove that the phase diagram of the 
effective hamiltonian $E_{\Lambda}(\smm)$ is rigorously given by the 
phase diagram of the truncated hamiltonian, except for domains of 
width $|U|^{-(k+1)}$ (which can be explicitely estimated) centered on the 
boundaries of the phase diagram of $H^{(k)}$. At this order nothing 
can be said concerning $(\mu_{e},\mu_{i})$ in these domains. Going to 
order $(k+1)$, the ground states of $H^{(k+1)}$ will give the ground 
states of $E_{\Lambda}(\smm)$ except for domains of width 
$|U|^{-(k+2)}$, and so on.

This method has been applied up to order $k=3$ to construct the phase 
diagrams for $d=1$ and 2, cubic and triangular lattices, both 
statistics, for systems with or without magnetic fields. However it 
is restricted by the conditions $\rho_{e}=\rho_{i}$ if $U<0$ and 
$\rho_{e}+\rho_{i}=1$ if $U>0$.

\paragraph{b) Expansion of $E_{\Lambda}(\sss)$ in powers of $|U|$} 
\hfill \\

This method has been used only in the canonical formalism. Contrary 
to the first method, it is valid for any rational densities 
$(\rho_{e},\rho_{i})$ of electrons and ions, but so far it is 
restricted to 1 dimension only.

Assuming $t_{xy} = t \neq 0$ only if $|x-y|=1$, then, using Rayleigh-Schr\"odinger perturbation theory, one obtains for the ground 
state energy density (for $U<0$)
\eq{
    e(\sss;\rho_{e}) = 2\rho_{e} - \frac{2t}{\pi} \sin \pi \rho_{e} - 
U \rho_{e}\rho_{i}
    	+ \frac{1}{4\pi t} \frac{|W_{q}|^{2}}{\sin\pi\rho_{e}} 
U^{2}\ln|U| + O(U^{2})
}
where $\rho_{e} = p/q$ (p prime with respect to q) and
\eq{
    W_{q}(\sss;\rho_{e}) = \frac{1}{q} \sum_{x=0}^{q-1} e^{-i 
2\pi\rho_{e}x} s_{x}.
}
Given $\rho_{e}=p/q$, the problem is then to find the configurations 
$\sss$ which minimize $e(\sss;\rho_{e})$.

\paragraph{c) Reduced phase diagrams} \hfill \\

This method is valid for arbitrary coupling constant $U$ and 
arbitrary densities $(\rho_{e},\rho_{i})$, resp. $(\mu_{e},\mu_{i})$, 
but gives only qualitative results. The idea is to select some 
restricted class of ions configurations, e.g. all periodic 
configurations with period $\leq 16$ together with mixtures of two 
such periodic configurations, and to search, by means of exact 
numerical computation, for the configurations in this class which 
minimize the energy. This yields the so called reduced phase diagram. 
It has been applied in 1 and 2 dimensions.

From these numerical calculations, one observes that this approach 
yields results consistent with those rigorously established in the 
limit of large or small $U$. Moreover the reduced phase diagram 
appears rather stable, \ie increasing the class of configurations 
considered, the boundaries of the previous reduced phase diagram 
become domains of smaller and smaller width, where new ground states 
configurations appear, while the main part of the diagram is not 
modified. The situation is similar to the case discussed above, 
passing from the truncated hamiltonian $H^{(k)}$ to $H^{(k+1)}$.

\paragraph{d) Finite systems} \hfill \\

In 1 and 2 dimensions explicit numerical computations of the energies 
have been developped to find the exact ground state configurations of 
some finite systems. In fact, it is this early approach which led to 
the conjecture of ``molecule formation'' and to the idea, later 
proved, that for small $|U|$ and small densities (e.g. 
$\rho_{e}=\rho_{i}<\frac{1}{4}$) the ground state is not periodic.

\section{Results for 1-dimensional systems}

In 1 dimension, using the closed loop expansion for the effective 
free energy, one first concludes that hard core bosons are identical 
to fermions \cite{2}.

In the following we consider only the attractive case $U<0$. From 
particle-hole symmetry one then obtains similar conclusion for $U>0$.

\newtheorem{theo}{Theorem}
Using the canonical formalism, the following results were obtained.
\begin{theo}[Strong coupling \cite{14}] \hfill \\ \vspace{-0.7cm}
\begin{list} {\alph{no}.}{\usecounter{no}}
\item Let $\rho_{e}=\rho_{i}=p/q$, with $p$ prime with respect to 
$q$, then for $|U|>U_{\text{cr}}(q)$ the ground state is the 
\emph{most homogeneous periodic configuration} with period $q$. The 
position of the ions on the sites $(0,1,\ldots,q-1)$ is given by 
$W_{x}=1$ for $x=k_{j}$ where $k_{j}$ is solution of the equation
\eq{	\label{1bis}
    p k_{j} = j \mod q \qquad j=0,1,\ldots,p-1.
}
\item Let $\rho_{i}=b \rho_{e}, b \neq 1$, then for 
$|U|>U_{\text{seg}}(b)$ the ground state is the \emph{segregated 
configuration} where all ions clump together.
\end{list}
\end{theo}

\newcommand{\re}{\rho_{e}}
\newcommand{\ri}{\rho_{i}}
\begin{theo}[Weak coupling \cite{15}] \hfill \\
 \vspace{-0.2 cm}
Given $\rho_{e}=p/q$ with $p$ prime with respect to $q$, and $|U|\ll 
q$, then
\begin{list} {\alph{no}.}{\usecounter{no}}
\item for $\rho_{i} \in ]\frac{p'}{q},\frac{p'+1}{q}[$, the ground 
state is a \emph{mixture} of two periodic configurations $\sss'$ and 
$\sss''$, with $\rho_{i}'=\frac{p'}{q}$, $\rho_{i}''=\frac{p'+1}{q}$ 
($\rho_{e}'=\rho_{e}''=p/q$), where the position of ions is given by 
eq. \eqref{1bis}, with $j=0,1,\ldots,p'-1$ (resp. $j=0,1,\ldots,p'$).
\item for $\rho_{i}=p_{i}/q$, $p_{i}$ not necessarily prime with 
respect to $q$,
\begin{enumerate}
\item if $\rho_{i}\in\, ]0.37,0.63[$, the ground state is the
\emph{periodic} configuration of period $q$, given by eq. \eqref{1bis}.
\item if $\rho_{i}<0.37$, or $\rho_{i}>0.63$, the ground state is a 
\emph{mixture} consisting of a periodic configuration, with period 
$q$ and ion density $\rho_{i}'=p_{i}'/q$, solution of eq. 
\eqref{1bis}, together with empty ($\rho_{i}''=0$) or full 
($\rho_{i}''=1$) configuration (and $\rho_{e}'=\rho_{e}''=p/q$), 
\emph{except} for a countable set of densities $(\re,\ri)$, with $\ri 
\in [\frac{1}{4},0.37] \cup [0.63,\frac{3}{4}]$, where the ground 
state is the periodic ground state, with period $q$, given by eq. 
\eqref{1bis}. This countable set is given by the solutions of an 
equation, and we have for example
\begin{alignat}{1} \notag
   \ri &= \re   = \frac{1}{4}, \frac{1}{3}, \frac{7}{20}, 
\frac{6}{17}, \frac{14}{39} \\
   \notag
   \ri &= 2\re =	\frac{1}{3}, \frac{6}{17},\frac{10}{18}, 
\frac{14}{39}, \ldots
\end{alignat}
\end{enumerate}
\end{list}
\end{theo}

This property shows that the periodicity of the pure phase is fixed 
by the electron density: it is the smallest period necessary to 
open a gap at the Fermi level. Let us also remark that the critical 
ion densities given above (\ie $\ri=0.37$ and $0.63$) are approximate 
values. The exact values are $\rho_{c}$ and $(1-\rho_{c})$, with 
$\rho_{c}$ solution of $2\pi\rho_{c} = \tan \pi \rho_{c}$.

The results of theorem 2 show that there is a close analogy with the Peierls instability; however this analogy is valid only for $\ri \in [\rho_{c},1-\rho_{c}]$ and can 
be seen as follows: For $U=0$ ($\re = p/q, \ri=p_{i}/q$ fixed) any 
ions configuration is a ground state and the probability to find an 
ion at a given site is uniform, equals to $\ri$. This is the 
``undistorted state'', which is metallic (no gap). For $U \neq 0$, 
sufficiently small, a particular configuration is selected which has 
period $q$. It corresponds to the ``distorted state'', which has a gap 
at the Fermi level and is insulating. For $\ri \not\in 
[\rho_{c},1-\rho_{c}]$ the ground state is a mixture of a metallic and 
insulating phase (except for exceptional densities) and such a 
situation does not occur in the standard theory of Peierls and 
Fr\"ohlich.

To extend these investigations to arbitrary values of the coupling 
constant $U$, and to avoid difficulties associated with mixtures, it is 
more convenient to work in the grand canonical formalism. Reduced 
phase diagrams were obtained in \cite{16} for several values of $U$, by 
means of exact numerical calculations. Some examples are illustrated 
on figure 1.

\begin{figure}[h]
\centering
\mbox{\subfigure[Strong coupling]{\epsfig{file=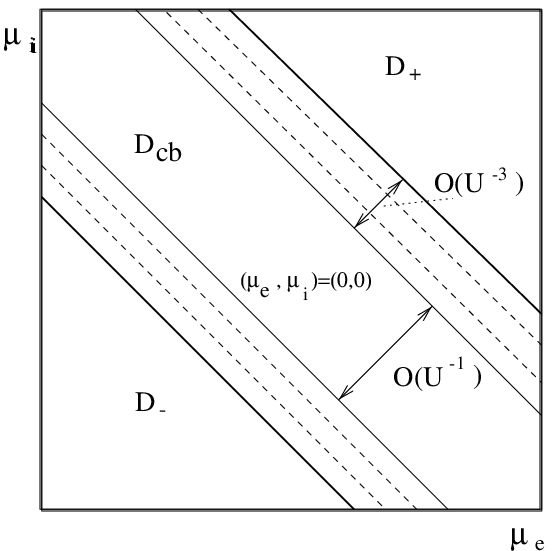}}}	\qquad
\mbox{\subfigure[Weak coupling]{\epsfig{file=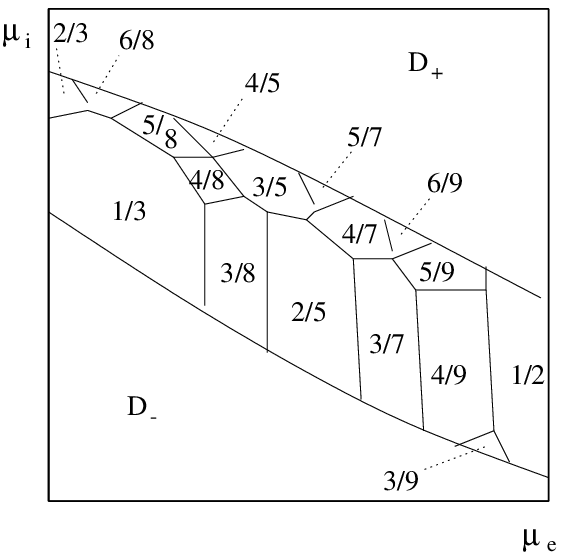}}} \\
   Figure 1
\end{figure}	\setcounter{figure}{1}
Figure 1a) shows part of the chemical potential plane for strong coupling, \ie $U<2t$. ${D}_{-}, {D}_{+}, {D}_{cb}$ represent domains where 
   the ground state is the empty ($\ri=0$) , the full ($\ri=1$), and 
   the chessboard configuration. The parallel stripes corresponds to 
   domains with $\re = \ri = \text{const}$.

Figure 1b) shows part of the chemical potential plane for weak coupling, \ie $U\in[-2t,0]$. The ``vertical'' stripes with same denominator 
   $q$ correspond to periodic ground states with fixed electron 
   density $\re=p/q$ and different ion densities $\ri=p_{i}/q$.

Combining all the results obtained so far, either rigorously or by reduced phase diagrams analysis, one is led to the following conclusion
\newtheorem{conj}{Conjecture} \renewcommand{\theconj}{}
\begin{conj}
Given the electron density $\re=p/q$, $p$ prime with respect to $q$, 
and the ion density $\ri=p_{i}/q$, $p_{i}$ not necessarily prime with 
respect to $q$, if the ground state is periodic then it has period $q$ 
and the position of the ions is given by the solutions of the equation
\begin{equation}   \label{1ter}
	p k_{j} = j \mod q, \qquad j=0,1,\ldots p_{i}-1.
\end{equation}
In all other cases, the ground state is either
\begin{enumerate}
  \item a mixture of two periodic configurations with $\ri = p_{i}/q$ 
  and $\ri = (p_{i}+1)/q$ given by the solutions of eq. \eqref{1ter}
  \item a mixture of one periodic configuration solution of eq. 
  \eqref{1ter}, together with either the empty $(\ri=0)$ or the 
  full ($\ri=1$) configuration.
  \item the segregated configuration.
\end{enumerate}
\end{conj}

Let us remark that eq. \eqref{1ter} is reminiscent of the circle map 
theorem, and appears in many situations.

\section{Results for two dimensional systems}

The ground state properties of the Falicov-Kimball model on the 
square lattice, in the limit of strong coupling, are discussed in the 
talk by T. Kennedy.

Reduced phase diagram analysis for the square lattice and arbitrary 
values of the coupling $U$ has been conducted by Watson and Lemanski 
\cite{17}, \cite{18}. Their results show that the properties discussed in sec. 2 
are not specific to one dimensional systems, but also appears (together 
with new properties) in two dimensions.

In this section we want to exhibit the difference between fermions 
and hard-core bosons, between square and triangular lattices, and to 
discuss the flux phase problem.

This analysis has been conducted, within the grand canonical formalism, 
in the limit of strong coupling, using the $|U|^{-1}$ expansion 
discussed in sec. 1.4. It is thus restricted to neutral (if $U<0$) or 
half-filled (if $U>0$) systems (ref. \cite{2}, \cite{19}).

The phase diagrams for fermions and hard-core bosons in the presence 
of an homogeneous magnetic field, defined by its flux $\Phi$ through 
elementary cells (in particular it could be zero), are represented on 
figures 2a and 2b for the square lattice, and on figures 4a and 4b for 
the triangular lattice. These phase diagrams are those of the 
truncated hamiltonians; the exact phase diagrams are identical except 
for small domains around the boundary curves which have been 
explicitely evaluated \cite{2}.

\begin{figure}[h] \setcounter{subfigure}{0}
\centering
\mbox{\subfigure[Phase diagram to order 3 for fermions on the square lattice] {
\raisebox{-3.2cm}{ }\hspace{6.5cm}} }
	\qquad
\mbox{\subfigure[Phase diagram to order 3 for hard-core bosons on the square lattice]{\raisebox{-3.2cm}{ }\hspace{6.5cm}}} \\
   Figure 2
\end{figure}	\setcounter{figure}{2}

\begin{center}
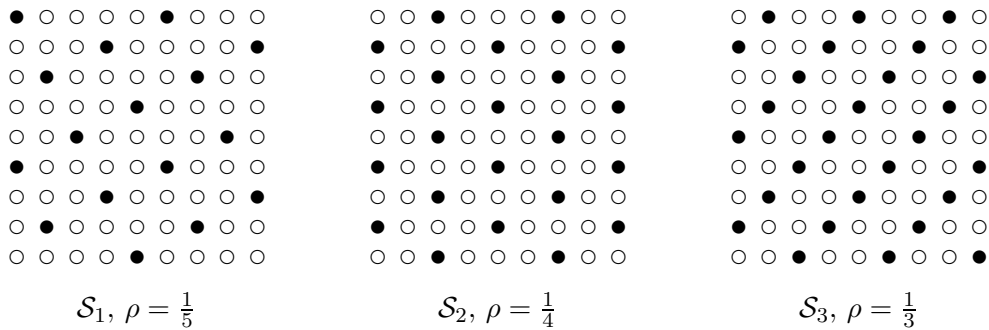
\begin{figure}[h]
   \setlength{\unitlength}{0.4cm}
\centering
\begin{picture}(32,11)(0,-2)
\put(0,0){\circle{0.4}} \put(1,0){\circle{0.4}} 
\put(2,0){\circle{0.4}} \put(3,0){\circle{0.4}} 
\put(4,0){\circle*{0.4}} \put(5,0){\circle{0.4}} 
\put(6,0){\circle{0.4}} \put(7,0){\circle{0.4}} 
\put(8,0){\circle{0.4}}
\put(0,1){\circle{0.4}} \put(1,1){\circle*{0.4}} 
\put(2,1){\circle{0.4}} \put(3,1){\circle{0.4}} 
\put(4,1){\circle{0.4}} \put(5,1){\circle{0.4}} 
\put(6,1){\circle*{0.4}} \put(7,1){\circle{0.4}} 
\put(8,1){\circle{0.4}}
\put(0,2){\circle{0.4}} \put(1,2){\circle{0.4}} 
\put(2,2){\circle{0.4}} \put(3,2){\circle*{0.4}} 
\put(4,2){\circle{0.4}} \put(5,2){\circle{0.4}} 
\put(6,2){\circle{0.4}} \put(7,2){\circle{0.4}} 
\put(8,2){\circle*{0.4}}
\put(0,3){\circle*{0.4}} \put(1,3){\circle{0.4}} 
\put(2,3){\circle{0.4}} \put(3,3){\circle{0.4}} 
\put(4,3){\circle{0.4}} \put(5,3){\circle*{0.4}} 
\put(6,3){\circle{0.4}} \put(7,3){\circle{0.4}} 
\put(8,3){\circle{0.4}}
\put(0,4){\circle{0.4}} \put(1,4){\circle{0.4}} 
\put(2,4){\circle*{0.4}} \put(3,4){\circle{0.4}} 
\put(4,4){\circle{0.4}} \put(5,4){\circle{0.4}} 
\put(6,4){\circle{0.4}} \put(7,4){\circle*{0.4}} 
\put(8,4){\circle{0.4}}
\put(0,5){\circle{0.4}} \put(1,5){\circle{0.4}} 
\put(2,5){\circle{0.4}} \put(3,5){\circle{0.4}} 
\put(4,5){\circle*{0.4}} \put(5,5){\circle{0.4}} 
\put(6,5){\circle{0.4}} \put(7,5){\circle{0.4}} 
\put(8,5){\circle{0.4}}
\put(0,6){\circle{0.4}} \put(1,6){\circle*{0.4}} 
\put(2,6){\circle{0.4}} \put(3,6){\circle{0.4}} 
\put(4,6){\circle{0.4}} \put(5,6){\circle{0.4}} 
\put(6,6){\circle*{0.4}} \put(7,6){\circle{0.4}} 
\put(8,6){\circle{0.4}}
\put(0,7){\circle{0.4}} \put(1,7){\circle{0.4}} 
\put(2,7){\circle{0.4}} \put(3,7){\circle*{0.4}} 
\put(4,7){\circle{0.4}} \put(5,7){\circle{0.4}} 
\put(6,7){\circle{0.4}} \put(7,7){\circle{0.4}} 
\put(8,7){\circle*{0.4}}
\put(0,8){\circle*{0.4}} \put(1,8){\circle{0.4}} 
\put(2,8){\circle{0.4}} \put(3,8){\circle{0.4}} 
\put(4,8){\circle{0.4}} \put(5,8){\circle*{0.4}} 
\put(6,8){\circle{0.4}} \put(7,8){\circle{0.4}} 
\put(8,8){\circle{0.4}}
\put(2,-2){${\cal S}_1$,  $\rho = {1 \over 5}$}
\put(12,0){\circle{0.4}} \put(13,0){\circle{0.4}} 
\put(14,0){\circle*{0.4}} \put(15,0){\circle{0.4}} 
\put(16,0){\circle{0.4}} \put(17,0){\circle{0.4}} 
\put(18,0){\circle*{0.4}} \put(19,0){\circle{0.4}} 
\put(20,0){\circle{0.4}}
\put(12,1){\circle*{0.4}} \put(13,1){\circle{0.4}} 
\put(14,1){\circle{0.4}} \put(15,1){\circle{0.4}} 
\put(16,1){\circle*{0.4}} \put(17,1){\circle{0.4}} 
\put(18,1){\circle{0.4}} \put(19,1){\circle{0.4}} 
\put(20,1){\circle*{0.4}}
\put(12,2){\circle{0.4}} \put(13,2){\circle{0.4}} 
\put(14,2){\circle*{0.4}} \put(15,2){\circle{0.4}} 
\put(16,2){\circle{0.4}} \put(17,2){\circle{0.4}} 
\put(18,2){\circle*{0.4}} \put(19,2){\circle{0.4}} 
\put(20,2){\circle{0.4}}
\put(12,3){\circle*{0.4}} \put(13,3){\circle{0.4}} 
\put(14,3){\circle{0.4}} \put(15,3){\circle{0.4}} 
\put(16,3){\circle*{0.4}} \put(17,3){\circle{0.4}} 
\put(18,3){\circle{0.4}} \put(19,3){\circle{0.4}} 
\put(20,3){\circle*{0.4}}
\put(12,4){\circle{0.4}} \put(13,4){\circle{0.4}} 
\put(14,4){\circle*{0.4}} \put(15,4){\circle{0.4}} 
\put(16,4){\circle{0.4}} \put(17,4){\circle{0.4}} 
\put(18,4){\circle*{0.4}} \put(19,4){\circle{0.4}} 
\put(20,4){\circle{0.4}}
\put(12,5){\circle*{0.4}} \put(13,5){\circle{0.4}} 
\put(14,5){\circle{0.4}} \put(15,5){\circle{0.4}} 
\put(16,5){\circle*{0.4}} \put(17,5){\circle{0.4}} 
\put(18,5){\circle{0.4}} \put(19,5){\circle{0.4}} 
\put(20,5){\circle*{0.4}}
\put(12,6){\circle{0.4}} \put(13,6){\circle{0.4}} 
\put(14,6){\circle*{0.4}} \put(15,6){\circle{0.4}} 
\put(16,6){\circle{0.4}} \put(17,6){\circle{0.4}} 
\put(18,6){\circle*{0.4}} \put(19,6){\circle{0.4}} 
\put(20,6){\circle{0.4}}
\put(12,7){\circle*{0.4}} \put(13,7){\circle{0.4}} 
\put(14,7){\circle{0.4}} \put(15,7){\circle{0.4}} 
\put(16,7){\circle*{0.4}} \put(17,7){\circle{0.4}} 
\put(18,7){\circle{0.4}} \put(19,7){\circle{0.4}} 
\put(20,7){\circle*{0.4}}
\put(12,8){\circle{0.4}} \put(13,8){\circle{0.4}} 
\put(14,8){\circle*{0.4}} \put(15,8){\circle{0.4}} 
\put(16,8){\circle{0.4}} \put(17,8){\circle{0.4}} 
\put(18,8){\circle*{0.4}} \put(19,8){\circle{0.4}} 
\put(20,8){\circle{0.4}}
\put(14,-2){${\cal S}_2$,  $\rho = {1 \over 4}$}
\put(24,0){\circle{0.4}} \put(25,0){\circle{0.4}} 
\put(26,0){\circle*{0.4}} \put(27,0){\circle{0.4}} 
\put(28,0){\circle{0.4}} \put(29,0){\circle*{0.4}} 
\put(30,0){\circle{0.4}} \put(31,0){\circle{0.4}} 
\put(32,0){\circle*{0.4}}
\put(24,1){\circle*{0.4}} \put(25,1){\circle{0.4}} 
\put(26,1){\circle{0.4}} \put(27,1){\circle*{0.4}} 
\put(28,1){\circle{0.4}} \put(29,1){\circle{0.4}} 
\put(30,1){\circle*{0.4}} \put(31,1){\circle{0.4}} 
\put(32,1){\circle{0.4}}
\put(24,2){\circle{0.4}} \put(25,2){\circle*{0.4}} 
\put(26,2){\circle{0.4}} \put(27,2){\circle{0.4}} 
\put(28,2){\circle*{0.4}} \put(29,2){\circle{0.4}} 
\put(30,2){\circle{0.4}} \put(31,2){\circle*{0.4}} 
\put(32,2){\circle{0.4}}
\put(24,3){\circle{0.4}} \put(25,3){\circle{0.4}} 
\put(26,3){\circle*{0.4}} \put(27,3){\circle{0.4}} 
\put(28,3){\circle{0.4}} \put(29,3){\circle*{0.4}} 
\put(30,3){\circle{0.4}} \put(31,3){\circle{0.4}} 
\put(32,3){\circle*{0.4}}
\put(24,4){\circle*{0.4}} \put(25,4){\circle{0.4}} 
\put(26,4){\circle{0.4}} \put(27,4){\circle*{0.4}} 
\put(28,4){\circle{0.4}} \put(29,4){\circle{0.4}} 
\put(30,4){\circle*{0.4}} \put(31,4){\circle{0.4}} 
\put(32,4){\circle{0.4}}
\put(24,5){\circle{0.4}} \put(25,5){\circle*{0.4}} 
\put(26,5){\circle{0.4}} \put(27,5){\circle{0.4}} 
\put(28,5){\circle*{0.4}} \put(29,5){\circle{0.4}} 
\put(30,5){\circle{0.4}} \put(31,5){\circle*{0.4}} 
\put(32,5){\circle{0.4}}
\put(24,6){\circle{0.4}} \put(25,6){\circle{0.4}} 
\put(26,6){\circle*{0.4}} \put(27,6){\circle{0.4}} 
\put(28,6){\circle{0.4}} \put(29,6){\circle*{0.4}} 
\put(30,6){\circle{0.4}} \put(31,6){\circle{0.4}} 
\put(32,6){\circle*{0.4}}
\put(24,7){\circle*{0.4}} \put(25,7){\circle{0.4}} 
\put(26,7){\circle{0.4}} \put(27,7){\circle*{0.4}} 
\put(28,7){\circle{0.4}} \put(29,7){\circle{0.4}} 
\put(30,7){\circle*{0.4}} \put(31,7){\circle{0.4}} 
\put(32,7){\circle{0.4}}
\put(24,8){\circle{0.4}} \put(25,8){\circle*{0.4}} 
\put(26,8){\circle{0.4}} \put(27,8){\circle{0.4}} 
\put(28,8){\circle*{0.4}} \put(29,8){\circle{0.4}} 
\put(30,8){\circle{0.4}} \put(31,8){\circle*{0.4}} 
\put(32,8){\circle{0.4}}
\put(26,-2){${\cal S}_3$,  $\rho = {1 \over 3}$}
\end{picture}
   \caption{\small Configurations $S_{1},S_{2}, S_{3}$ which appears in fig. 2 (a) and (b). $\bar{S}_{1}, \bar{S}_{2}, \bar{S}_{3}$ are obtained by 
   particle-hole transformation. $S_{-}, S_{+}, S_{\text{cb}}$ are 
   respectively the empty ($\rho=0$), the full ($\rho=1$), and the 
   chessboard ($\rho=\frac{1}{2}$) configurations.}
\end{figure}
\end{center}

\begin{figure}[h]	
\centering
\mbox{\subfigure[Phase diagram to order 2 for fermions on the triangular 
   lattice.]
	{\raisebox{-3.2cm}{ }\hspace{6.5cm}}}\qquad
\mbox{\subfigure[Phase diagram to order 2 for hard-core bosons on the triangular 
   lattice.]{\raisebox{-3.2cm}{ }\hspace{6.5cm}}} \\
   Figure 4
\end{figure}	\setcounter{figure}{4} 

In fig. 4 (a) and (b), $\tau_{-}$ and $\tau_{+}$ are the empty ($\rho=0$) 
and full ($\rho=1$) configurations; $\tau_{5}$ and $\bar{\tau}_{5}$ 
are the periodic configurations with densities $\frac{1}{3}$ and 
$\frac{2}{3}$.

The analysis of the triangular lattice has been extended to order 3. 
At this order the phase diagramms exhibit periodic configurations with 
densities $\rho = (0,\frac{1}{7},\frac{1}{5},\frac{1}{4},\frac{1}{4},
\frac{1}{3},\frac{2}{5},\frac{4}{9}, \frac{1}{2},\frac{1}{2})$ as well 
as $(1-\rho)$. We note that there exist two different structures with 
densities $\frac{1}{4}$, and with density $\frac{1}{2}$. The 
interested reader should consult the original article \cite{2}.

It remains to discuss the flux phase problem. In particular one would 
like to know in what cases, if any, the magnetic flux will decrease 
the energy of the system, and what is the optimal magnetic flux to 
obtain the state of minimum energy, \ie the ground state.

For hard-core bosons it is easy to see that the optimal magnetic flux is always 
zero (diamagnetic inequality) \cite{2}.

For fermions the situation is more subtle; it depends on the lattice 
and the densities. On the square lattice, one finds that for 
density $\frac{1}{2}$ the optimal flux is $\phi = \pi$ through each 
plaquette, and the ion configuration is the chessboard structure; for 
densities $\frac{1}{3}$ and $\frac{2}{3}$ the optimal flux is no 
longer uniform but periodic with period 3, and $\phi=0$ or $\pi$ (the 
ion configurations are the structures $S_{3}, \bar{S}_{3}$ of figure 
5); similarly for densities $\frac{1}{5}$ and $\frac{4}{5}$, at order 3 the optimal flux is 
non uniform, but periodic with $\phi=0$ or arbitrary (and the configurations are $S_{1}, \bar{S}_{1}$); for densities 
$0$ and $1$, the fluxes are arbitrary. We note that at this order the 
configurations with densities $\frac{1}{4}$ and $\frac{3}{4}$ do not 
appear (but they might appear at the next order however). We also 
should remark that on the square lattice the optimal flux is 
maximum ($\ie \phi=\pi$ everywhere) for density $\rho=\frac{1}{2}$, which is 
the maximum possible density, because of the particle-hole symmetry.

The situation is very different for the triangular lattice. One finds 
that the optimal flux is maximum ($\phi = \pi$ everywhere) for 
densities $\rho = \frac{1}{4}, \frac{1}{3}, \frac{2}{3}, \frac{3}{4}$; 
however for density $\rho = \frac{2}{5}, \frac{1}{2}, \frac{3}{5}$, 
the optimal flux is not uniform, but periodic, with $\phi=0$ or 
$\pi$. (Let us recall that on the triangular lattice the particle-hole 
transformation is not a symmetry). Similarly for densities 
$\rho=\frac{1}{7}$ and $\frac{5}{7}$, the optimal flux is not uniform, 
but periodic, with $\phi=0$ or arbitrary (this arbitrariness may be 
lifted at the next order); finally for densities $0$ and $1$, the 
fluxes are arbitrary.

The results for the flux phase problem are illustrated on figure 5.
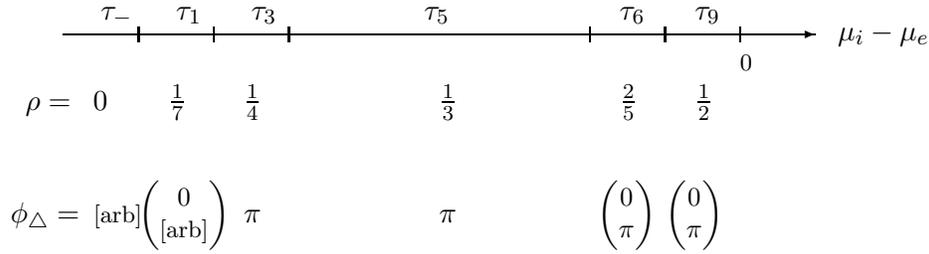
\begin{figure}[h]
\begin{center}
\unitlength 1cm
\begin{picture}(18,5)(0,1)
   \put(0,5){\underline{Square Lattice}}
   \put(1,4){\vector(1,0){10}}	\put(11.3,3.92){$\mu_{i}-\mu_{e}$}
   \put(2,4.1){\line(0,-1){0.2}}	\put(1.5,4.2){$S_{-}$}
   \put(3,4.1){\line(0,-1){0.2}}	\put(2.5,4.2){$S_{1}$}
   \put(4,4.1){\line(0,-1){0.2}}	\put(3.5,4.2){$S_{3}$}
   \put(6,4.1){\line(0,-1){0.2}}	\put(5.8,4.2){$S_{4}$}  \put(5.95,3.52){\footnotesize 0}
   \put(8,4.1){\line(0,-1){0.2}}	\put(8.4,4.2){$\bar{S}_{3}$}
   \put(9,4.1){\line(0,-1){0.2}}	\put(9.4,4.2){$\bar{S}_{1}$}
   \put(10,4.1){\line(0,-1){0.2}}	\put(10.4,4.2){$\bar{S}_{+}$}
   \put(0.5,3){$\rho =$}
	\put(1.4,3){$0$}
	\put(2.4,3){$\frac{1}{5}$}
	\put(3.4,3){$\frac{1}{3}$}
	\put(6,3){$\frac{1}{2}$}
	\put(8.4,3){$\frac{2}{3}$}
	\put(9.4,3){$\frac{4}{5}$}
	\put(10.4,3){$1$}
   \put(0.3,1.5){$\phi_{\Box}=$}
	\put(1.4,1.5){\footnotesize [arb]}
	\put(2,1.5){$\small \begin{pmatrix} 0 \\ \text{\footnotesize [arb]} \end{pmatrix}$}
	\put(3.1,1.5){$\small \begin{pmatrix} 0 \\ \pi \end{pmatrix}$}
	\put(6,1.5){$\pi$}
	\put(8.1,1.5){$\small \begin{pmatrix} 0 \\ \pi \end{pmatrix}$}
	\put(9,1.5){$\small \begin{pmatrix} 0 \\ \text{\footnotesize [arb]} \end{pmatrix}$}
	\put(10.4,1.5){\footnotesize arb}
\end{picture}
\begin{picture}(18,5)(0,1)
   \put(0,5){\underline{Triangular Lattice}}
   \put(1,4){\vector(1,0){10}}	\put(11.3,3.92){$\mu_{i}-\mu_{e}$}
   \put(2,4.1){\line(0,-1){0.2}}	\put(1.5,4.2){$\tau_{-}$}
   \put(3,4.1){\line(0,-1){0.2}}	\put(2.5,4.2){$\tau_{1}$}
   \put(4,4.1){\line(0,-1){0.2}}	\put(3.5,4.2){$\tau_{3}$}
   				\put(5.8,4.2){$\tau_{5}$}
   \put(8,4.1){\line(0,-1){0.2}}	\put(8.4,4.2){$\tau_{6}$}
   \put(9,4.1){\line(0,-1){0.2}}	\put(9.4,4.2){$\tau_{9}$}
   \put(10,4.1){\line(0,-1){0.2}}   \put(10,3.52){\footnotesize 0}
   \put(0.5,3){$\rho =$}
	\put(1.4,3){$0$}
	\put(2.4,3){$\frac{1}{7}$}
	\put(3.4,3){$\frac{1}{4}$}
	\put(6,3){$\frac{1}{3}$}
	\put(8.4,3){$\frac{2}{5}$}
	\put(9.4,3){$\frac{1}{2}$}
   \put(0.3,1.5){$\phi_{\triangle}=$}
	\put(1.4,1.5){\footnotesize [arb]}
	\put(2,1.5){$\small \begin{pmatrix} 0 \\ \text{\footnotesize [arb]} \end{pmatrix}$}
	\put(3.4,1.5){$\pi$}
	\put(6,1.5){$\pi$}
	\put(8.1,1.5){$\small \begin{pmatrix} 0 \\ \pi \end{pmatrix}$}
	\put(9,1.5){$\small \begin{pmatrix} 0 \\ \pi \end{pmatrix}$}
\end{picture}
   \caption{Optimal magnetic fluxes, [arb] means arbitrary at order 3, $\ri=\re=\rho.$}
\end{center}
\end{figure}

We conclude this discussion with the following theorem \cite{2}. \\ \vspace{1cm}

\begin{theo}
For fermions systems
\begin{enumerate}
\item for the configurations $s_{+}=\set{s_{x}=+1}$ and $s_{-} = 
\set{s_{x}=-1}$, the effective hamiltonian is independant of the 
magnetic fluxes.
\item for any configuration $\sss \neq s_{+}$ and $s_{-}$, there 
exists $U_{0}(\sss)$ such that for $U \geq U_{0}(\sss)$, the optimal 
fluxes (\ie those wich minimize the energy) are
\begin{enumerate}
\item for the square lattice
\begin{equation}
	\phi_{\Box}^{\text{min}} (\sss) = 
	\begin{cases}
	\pi & \text{if 2 sites of $\Box$ are occupied} \\
	0   & \text{otherwise}
	\end{cases}
\end{equation}
\item for the triangular lattice
\begin{equation}
   \phi_{\triangle}^{\text{min}} =
   \begin{cases}
	\pi & \text{if 0 or 1 site of $\triangle$ are occupied} \\
	0   & \text{otherwise}
   \end{cases}
\end{equation}
\end{enumerate}
\end{enumerate}
\end{theo}

\section{Preliminary results for $\infty$-dimension Bethe lattice system}

In this section, the system is defined on the Bethe lattice with 
coordination number $z$ (fig.~6) and by the hamiltonian
\begin{equation}
	H(\vect{W}) = - \sum_{x,y}t_{xy}\, \acroix_{x}a_{y} +
	U \sum_{x}W_{x}\acroix_{x}a_{x}
\end{equation}
with $W_{x} \in \set{0,1}$ and
\begin{equation}
	t_{xy}=t_{yx} =
	\begin{cases}
	  \frac{t}{\sqrt{z}}   & \text{if $|x-y|=1$} \\
	  0                    & \text{otherwise}
	\end{cases}
\end{equation}

We shall then consider the limit $z\tend\infty$. The Falicov-Kimball 
model on this infinite dimensional Bethe lattice has attracted a 
considerable interest since 1989 and the reader should consult the 
references \cite{22}, \cite{23} for more informations.

\begin{figure}[h]
\begin{center}
   \epsfig{file=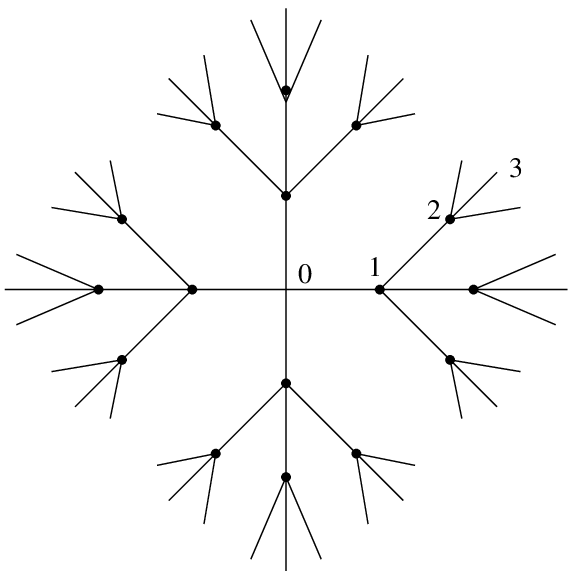}
   \caption{Bethe lattice. The shells $l=0,1,2,3$ are 
   represented.}
\end{center}
\end{figure}

The density of electrons on the site $x$ is expressed as
\begin{equation}
	\left< \acroix_{x}a_{x} \right>(\beta,\mu_{e},\mu_{i}) =
	\frac{1}{\beta} \sum_{n} G_{nn}(\omega_{n})
\end{equation}
with $\omega_{n}=\frac{\pi}{\beta}(2n+1)$ the Matsubara frequencies.

\newcommand{\bracket}[1]{\left< #1 \right>}
The strategy proceeds as follows. Using
\begin{enumerate}
\item Grassmann variables
\item the fact that $H$ is quadratic in $\acroix_{x}, a_{x}$
\item the fact that there are no closed loop on Bethe lattice \\
and introducing the complex variable
\begin{equation}
	\xi_{n}=i \omega_{n} - \mu_{e}
\end{equation}
we obtain the explicit expression
\begin{equation}	\label{40}
	G_{xx}(\omega_{n}) = \frac{\bracket{W_{x}}-1}{\xi_{n} + 
	I_{x}(\omega_{n})} - \frac{\bracket{W_{x}}}{\xi_{n} + U + 
	I_{x}(\omega_{n})}
\end{equation}
where
\begin{equation}	\label{41}
	I_{x}(\omega_{n}) = \frac{1}{2} \sum_{y: |x-y|=1} t^{2} 
	G_{yy}^{\Lambda\setminus x}(\omega_{n})
\end{equation}
($\Lambda\setminus x$ is the Bethe lattice with the site $x$ deleted) 
and
\begin{gather}
  \bracket{W_{x}} = \left[ 1+ e^{-\beta(\mu_{i}-Y_{x}(\beta,\mu_{e}))} 
  \right]^{-1}\\
  Y_{x}(\beta,\mu_{e}) = \frac{1}{\beta} 
  \sum_{n}\set{\ln[\xi_{n}+I_{x}(\omega_{n})] - \ln[\xi_{n} + U + 
  I_{x}(\omega_{n})]}
\end{gather}
\item At this point one takes the thermodynamic limit in a symmetrical 
manner, by increasing to infinity the number of shells. In this limit
\begin{equation}
	G_{xx}(\omega_{n}) = G_{l}(\omega_{n}), \qquad \bracket{W_{x}} = 
	\alpha_{l}
\end{equation}
where $l=0,1,2,\ldots$ denotes the shells ($l=0$ is the center 
point, $l=1$ are the $z$ nearest neighbours, and so on).
\item Taking the limit of infinite dimension, $z\tend\infty$, we 
have from \eqref{41} and \eqref{40}
\begin{gather}
	I_{x}(\omega_{n}) = I_{l}(\omega_{n}) = t^{2}G_{l+1}(\omega_{n}) \\
\label{(5)}
	G_{l}(\omega_{n}) = 
	\frac{\alpha_{l}-1}{\xi_{n}+t^{2}G_{l+1}(\omega_{n})} - 
	\frac{\alpha_{l}}{\xi_{n}+U+t^{2}G_{l+1}(\omega_{n})} \\
\label{47}
	\alpha_{l} = \left[   1+e^{-\beta(\mu_{i}-Y_{x}(\beta,\mu_{e}))} \right]^{-1}
\end{gather}
\item We can then take the zero temperature limit $\beta\tend\infty$, to obtain
\eq{	\label{(6)}
    Y_{l}(\mu_{e}) = \frac{1}{2\pi} \int_{-\infty}^{\infty}\romd \omega\,
    \set{\ln[\xi + t^{2}G_{l+1}(\omega)] - \ln[\xi + U + t^{2}G_{l+1}(\omega)]}
}
\item We thus conclude from \eqref{47} that in the limit $\beta\tend\infty$
\begin{align}	\label{(7)}
	\text{either\ } &\alpha_{l}=0 \quad \text{and this happens iff\ } 
	Y_{l}(\mu_{e}) > \mu_{i} \\
	\text{or\ } &\alpha_{l}=1 \quad \text{and this happens iff\ }
	Y_{l}(\mu_{e}) < \mu_{i}
\end{align}
\item The problem is reduced to the following ones: \\
Given $\vect{\alpha}=[\alpha_{0},
\alpha_{1},\ldots,\alpha_{p-1}], \alpha_{\rho} \in \set{0,1}$,
\begin{enumerate}
 \item show that eq. \eqref{(5)} has a unique solution, 
 $G_{l}(\omega_{n}), l=0,1,\ldots,n-1$
 \item from $G_{l}(\omega_{n})$ and eq. \eqref{(6)} determines $Y_{l}(\mu_{e})$
 \item Using eq. \eqref{(7)} find those $(\mu_{e},\mu_{i})$ for which 
 the ``periodic'' configuration $\vect{\alpha}$ (periodic with respect 
 to successive shells) is the ground state.
\end{enumerate}
\end{enumerate}
With this procedure, the following result was obtained (\cite{20}, \cite{21}).
\begin{theo}
For $\mu_{e} \in ]2t,U-2t[$, then
\begin{itemize}
\item for $\mu_{i}-\mu_{e} < -t^{2}/U$, the ground state is the empty configuration ($W_{x}=0$).
\item for $\mu_{i}-\mu_{e} > t^{2}/U$, the ground state is the full configuration ($W_{x}=1$).
\item for $\mu_{i}-\mu_{e} \in [-A,A]$, with
\eq{
    A = \frac{t^{2}}{U} -4\frac{t^{4}}{U^{3}} + 27\frac{t^{6}}{U^{5}} + O(\frac{t^{8}}{U^{7}})
}
the ground states are the period 2 configurations $\vect{\alpha}=[0,1]$, and $\vect{\alpha}=[1,0]$.
\item for $\mu_{i}-\mu_{e} \in [-B,-A]$, and for $\mu_{i}-\mu_{e} \in [A,B]$ with
\eq{
    B = \frac{t^{2}}{U} -6\frac{t^{6}}{U^{5}} + O(\frac{t^{8}}{U^{7}})
}
the ground states are the period 3 configurations given respectively by $\vect{\alpha}=$ $[0,0,1],$ $ [0,1,0],$ $ [1,0,0]$ and by $\vect{\alpha}=$$[0,1,1],$$ [1,0,1], $$[1,1,0]$.
\end{itemize}
\end{theo}

\section{Conclusion}

Although extremly simple in its formulation, the Falicov-Kimball models exhibit a very rich structure in all dimension. In particular, on can prove
\begin{itemize}
\item Periodic ground states
\item Molecules formation
\item Segregation
\item Coexistence of phases
\item Metal-insulator transition
\item Peierls instability
\item Devil staircase (maybe complete)
\item Farey tree properties
\item Flux phases
\end{itemize}

Among the open problems, on can mention the following
\begin{itemize}
\item Prove the conjecture for 1-dimensional systems
\item Find the exact equation for the segregated configuration (1-dim)
\item Find the conditions for the ground state to be periodic (in any dimension)
\item Prove the devil staircase structure
\item Find the low temperature properties
\item Find the thermodynamic properties of the general class of system consisting of quantum particles interacting with classical fields.
\end{itemize}

\raisebox{-3cm}[0cm][0cm]{\parbox{\textwidth}{A large number of references concerning the Falicov-Kimball model in 1 and 2 dimensions can be found in [1] and [2] below and will not be reproduced here, unless quoted in the text.}}

\bibliographystyle{Vincent}

\begin{small}
    \bibliography{Falicov}

\newcommand{\SortNoop}[1]{}
\begin{thebibliography}{10} \vspace{2.5cm}

\bibitem{1}
Ch. Gruber and N.~Macris.
\newblock The Falicov-Kimball model: a review of exact results and extensions.
\newblock {\em Helv. Phys. Acta {\bf69}\/} (1996), p.  850.

\bibitem{2}
Ch. Gruber N. Macris~A. Messager and D.~Ueltschi.
\newblock Ground states and flux configurations of the two-dimensional
  Falicov-Kimball model.
\newblock {\em J. Stat. Phys. {\bf86}\/} (1997), p. ~57.

\bibitem{3}
Ch. Gruber and D.~Ueltschi.
\newblock {\em Physica A {\bf232}\/} (1996), p.  616.

\bibitem{4}
T.~Kennedy and E.~Lieb.
\newblock {\em Physica A {\bf138}\/} (1986), p.  320.

\bibitem{5}
L.~M. Falicov and J.~C. Kimball.
\newblock {\em Phys. Rev. Lett. {\bf22}\/} (1969), p.  957.

\bibitem{6}
J.~L. Lebowitz and N.~Macris.
\newblock {\em Rev. Math. Phys. {\bf6}\/} (1994), p.  927.

\bibitem{7}
C.~Borgs~R. Kotecky and D.~Ueltschi.
\newblock {\em Comm. Math. Phys. {\bf181}\/} (1996), p.  409.

\bibitem{8}
R.~Kotecky and D.~Ueltschi.
\newblock Effective interactions due to quantum fluctuations.
\newblock {\em Preprint\/} (1998).

\bibitem{9}
D.~Ueltschi.
\newblock Discontinuous phase transitions in quantum lattice systems.
\newblock {\em PhD thesis EPFL\/} (1998).

\bibitem{10}
N.~Datta~R. Fernandez and J.~Fr\"ohlich.
\newblock {\em J. Stat. Phys. {\bf84}\/} (1996), p.  455.

\bibitem{11}
N.~Datta R. Fernandez~J. Fr\"ohlich and L.~Rey-Bellet.
\newblock {\em Helv. Phys. Acta {\bf69}\/} (1996), p.  752.

\bibitem{12}
A.~Messager and S.~Miracle-Sol\'e.
\newblock {\em Rev. Math. Phys. {\bf8}\/} (1996), p.  271.

\bibitem{13}
S.~Miracle-Sol\'e.
\newblock {\em Physica A {\bf232}\/} (1996), p.  686.

\bibitem{14}
P.~Lemberger.
\newblock {\em J. Phys. A {\bf25}\/} (1992), p.  715.

\bibitem{15}
J.~K. Freericks~Ch. Gruber and N.~Macris.
\newblock {\em Phys. Rev. B {\bf53}\/} (1996), p.  16189.

\bibitem{16}
Ch. Gruber~D. Ueltschi and J.~J\c{e}drzejewski.
\newblock {\em J. Stat. Phys. {\bf76}\/} (1994), p.  125.

\bibitem{17}
G.~I. Watson and R.~Lemanski.
\newblock {\em J. Physics {\bf7}\/} (1995), p.  9521.

\bibitem{18}
G.~I. Watson.
\newblock {\em Physica A {\bf246}\/} (1997), p.  253.

\bibitem{19}
Ch. Gruber J. J\c{e}drzejewski~P. Zemberger.
\newblock {\em J. Stat. Phys. {\bf66}\/} (1992), p.  913.

\bibitem{20}
Ph. Royer.
\newblock Mod\`ele de Falicov-Kimball sur le r\'eseau de Bethe avec nombre de
  coordination infini.
\newblock {\em Travail pratique de dipl\^ome EPFL (unpublished)\/} (1998).

\bibitem{21}
J.~K. Freericks~Ch. Gruber and N.~Macris.
\newblock Phase separation and the segregation principle in the infinite $U$
  spinless Falicov-Kimball model.
\newblock {\em (to be published)\/}.

\bibitem{22}
P.~G.~J. van Dongen.
\newblock {\em Phys. Rev. B {\bf45}\/} (1992), p.  2267.

\bibitem{23}
A.~George and G.~Kotliar.
\newblock {\em Phys. Mod. Phys. {\bf68}\/} (1996), p. ~14.

\end{thebibliography}
\end{small}

\end{document}